\pdfoutput=1
\documentclass[nopubldata]{lpor2014}
\graphicspath{{figs/}}

\begin{document}
 \titlefigure{Cover}

% Give an abstract text:
\abstract{We report on efficient nonlinear generation of ultrafast, higher order ``perfect'' vortices at the green wavelength. Based on Fourier transformation of the higher order Bessel-Gauss beam generated through the combination of spiral phase plate and axicon we have transformed the Gaussian beam of the ultrafast Yb-fiber laser at 1060 nm into perfect vortices of power 4.4 W and order up to 6. Using single-pass second harmonic generation (SHG) of such vortices in 5-mm long chirped MgO-doped, periodically poled congruent LiNbO$_3$ crystal we have generated perfect vortices at green wavelength with output power of 1.2 W and vortex order up to 12 at single-pass conversion efficiency of 27\% independent of its order. This is the highest single-pass SHG efficiency of any optical beams other than Gaussian beams. Unlike the disintegration of higher order vortices in birefringent crystals, here, the use of quasi-phase matching process enables generation of high quality vortices even at higher orders. The green perfect vortices of all orders have temporal and spectral width of 507 fs and 1.9 nm, respectively corresponding to a time-bandwidth product of 1.02.}
\title{Efficient nonlinear generation of high power, higher order, ultrafast ``perfect'' vortices in green }
\titlerunning{Nonlinear generation of ``perfect'' vortices}
\author{N. Apurv Chaitanya \inst{1,2,*}, M. V. Jabir\inst{1}, G. K. Samanta\inst{1}}

\authorrunning{N. Apurv Chaitanya et al}

\institute{
Photonic Sciences Lab., Physical Research Laboratory, Navarangpura, Ahmedabad 380009, Gujarat, India
\and
Indian Institute of Technology-Gandhinagar, Ahmedabad 382424, Gujarat, India
}
\mail{\email{apurv@prl.res.in}}

\maketitle
Optical vortices, having phase singularities (phase dislocations) in the wavefront, carry vanishing intensity at the singular point. Due to the screw-like (helical) phase structure around the point of singularity, such beams carry orbital angular momentum (OAM). The phase distribution of the optical vortices can be represented as $\sim\exp(il\theta)$, with $\theta$ as azimuthal angle and the integer $l$, as topological charge (order).  Each photon of the beam carries OAM of $l\hslash$. Since the discovery of OAM associated with optical vortices \cite{1}, these beams have drawn a great deal of attention from various fields of science and technology including high resolution microscopy \cite{2}, quantum information \cite{3}, material processing \cite{4} and particle micro-manipulation and lithography \cite{5}. However, the major challenge through these applications is the need for sources of coherent radiation in vortex spatial profiles at different topological charges and wavelengths.

 The optical vortices are typically generated by spatial phase modulation of Gaussian beams using different types of modulators including spatial light modulators (SLMs) and spiral phase plates (SPP). However, all of those modulators have its own advantages and disadvantages in terms of wavelength coverage, mode conversion efficiency, damage threshold and power handling capabilities and cost \cite{6}. On the other hand, nonlinear frequency conversion techniques can be used to access high power/energy optical vortices at different wavelengths across the electromagnetic spectrum inaccessible to lasers. As such, frequency up-conversion of high power, ultrafast, optical vortices at 1.064 $\mu$m has given access to higher order optical vortices ($l=12$) in the green at 0.532 $\mu$m \cite{6}. Similarly, frequency down-conversion in optical parametric oscillators has produced optical vortex beam of order, $l=1$, with spectral tunability across 1 $\mu$m \cite{7} and 2 $\mu$m \cite{8}. So far, the nonlinear generation of vortex beams with high power/energy has been restricted to the vortices of lower orders \cite{6}. In case of optical vortices, the beam area  and  divergence \cite{Reddy_Vortex_Divergence} increases linearly with the order of the vortices. As a result, the nonlinear parametric gain, which depends on the intensity of the driving fields and the overlapping integral of the interacting beams, and thus the conversion efficiency of the nonlinear processes decreases with the order of the vortices. Therefore, nonlinear generation of higher order optical vortices especially with high power/energy requires optical vortices with beam area and beam divergence independent to their orders. Fortunately, recent development in the field of structured beams provided a new class of optical vortex beams known as ``perfect'' vortex \cite{Praveen_OL,Jabir}. Unlike vortices, the perfect vortices have annular ring radius independent of its orders.

             Typically, Fourier transformation of the Bessel-Gauss (BG) beam of different orders is used to generate perfect vortices \cite{Praveen_OL}. However, the combination of SPP and axicon converting the Gaussian beam into Laguerre-Gaussian (LG) beams and BG beams of different orders, as presented in the current report, can be considered as one of the simplest schemes for generating high power perfect vortices of different orders. The complex field amplitude of the experimentally realizable perfect vortex of order, \textit{l} at the back focal plane ($z=0$) of the Fourier transforming lens may be represented in polar coordinates as \cite{Praveen_OL} 

\begin{equation}
E(\rho,\theta)=i^{l-1}\frac{w_g}{w_o} \exp\left( -\frac{(\rho-\rho_r)^2}{w_o^2}\right) \exp\left(il\theta\right)  
\label{Field}
\end{equation}

Here, $w_g$ is the waist radius of the Gaussian beam confining the vortex beam. $\rho_r=f\sin(n-1)\alpha$  is the radius of the perfect vortex ring, $f$ is the focal length of the Fourier transforming lens, and $n$ and $\alpha$ are respectively the refractive index and base angle of the axicon. $2w_o$ ($w_o =2f/kw_g$, the Gaussian beam waist radius at the focus) is the annular width of the perfect vortex. $k=2\pi/\lambda$ is the wave vector of the beam of wavelength $\lambda$ in free space. Using Eqn. (\ref{Field}), we can write the intensity of the perfect vortex as,                     
\begin{equation}
I(\rho,\theta)=\left(\frac{w_g}{w_o}\right)^2 \exp\left( -2\frac{(\rho-\rho_r)^2}{w_o^2}\right) 
\label{Intensity}
\end{equation}
As evident from the Eqn. (\ref{Intensity}), the intensity of the perfect vortex is independent of its order. Therefore, one can expect the nonlinear frequency conversion efficiency of such perfect vortices to be independent of their orders \cite{Noncollinear_oam}. Using such perfect vortices, here, we report, for the best of our knowledge, the first experimental demonstration of nonlinear generation of perfect vortices of power $>1.2$ W at green with conversion efficiency as high as 27 \% and topological charge (order) as high as 12. We have also experimentally verified that at high power regime, the conversion efficiency of perfect vortices is independent to its order.

\begin{figure}
\centering%\sidecaption
\includegraphics*[width=\columnwidth ]{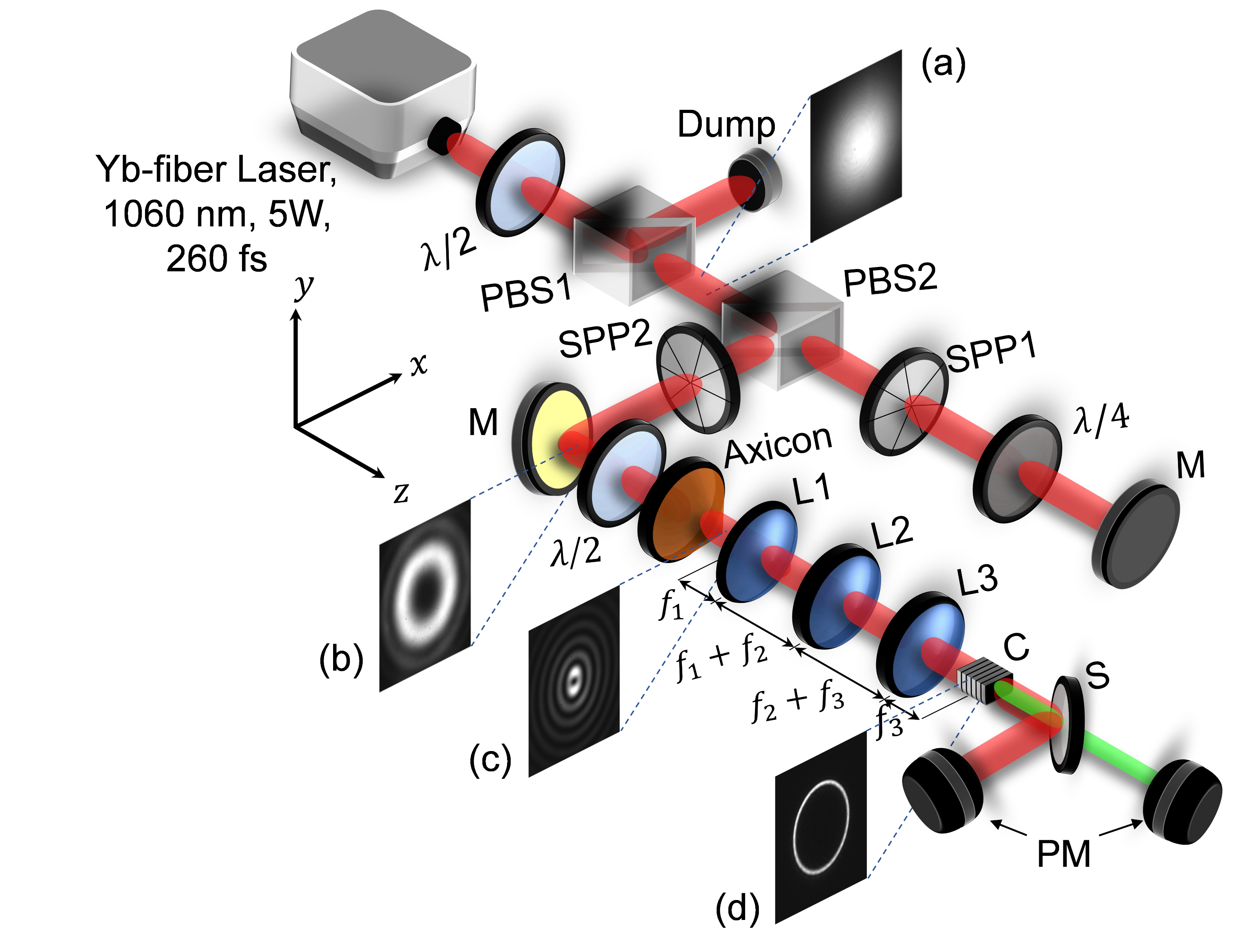}
\caption{Schematic of the experimental setup for nonlinear generation of ultrafast perfect vortices. $\lambda/2$; half-wave plate, PBS1,2; polarizing beam splitter cube, SPP1,2; spiral phase plates, $\lambda/4$; quater-wave plate, L1-3; lens, M; mirrors, C; MgO:CLN crystal for frequency-doubling, S; Dichroic mirror; PM, Power meter. (a-d) Spatial intensity profile of the beams recorded at different positions along the propagation direction.}
\label{Fig1}
\end{figure}

The schematic of the experimental set up is shown in Fig.\ref{Fig1}. An ultrafast Yb-fiber laser at 1060 nm similar to Ref. \cite{6} producing output in Gaussian (Fig. \ref{Fig1}(a)) intensity distribution ($M^2 < 1.1$) is used as the pump source. The laser output has temporal and spectral width of 260 fs and 15 nm respectively at a repetition rate of 78 MHz. The input power to the nonlinear crystal is controlled using a half wave plate ($\lambda/2$) and polarizing beam splitter cube (PBS1). Using only two spiral phase plates, SPP1 and SPP2, of winding numbers 1 and 2 respectively and a vortex – doubler \cite{6} comprising of PBS2, quarter wave plate ($\lambda/4$) and a plane mirror (M) with high reflectance for 1060 nm, we have generated optical vortices of order $l_p=1-6$.  The intensity profile of the generated vortex beam ($l_p=2$) is given in Fig. \ref{Fig1}(b). The antireflection (AR) coated axicon of apex angle $\sim178^{\circ}$, converts vortex beam (LG beam) into BG beam (see Fig. \ref{Fig1} (c)) of same order.  A plano-convex lens, L1, of focal length, $f_1 =25$ mm Fourier transforms the BG beam into perfect vortices. The imaging system comprising two plano-convex lenses L2 and L3 with focal lengths $f_2=200$ mm and $f_3=50$ mm respectively, images the perfect vortex (see Fig. \ref{Fig1}(d)) at the centre of the nonlinear crystal to a measured beam radius of $ \sim166~\mu$m.  A 5-mm-long and 2 x 1 mm$^2$ in aperture, MgO-doped, periodically poled congruent LiNbO$_3$ (MgO:CLN) crystal(C) with linear chirped grating period of 6.61-6.91 $\mu$m is used for second harmonic generation (SHG) of the pump vortex beam. The crystal has spectral acceptance bandwidth of 15 nm to cover entire pump spectrum   and phase-matching temperature of $130^{\circ}\mathrm{C}$ to avoid any detrimental photorefractive effect. Both the faces of the crystal is AR coated for 530 and 1060 nm. The crystal is housed in an oven whose temperature can be varied up to $200^{\circ}\mathrm{C}$  in steps of $0.1^{\circ}\mathrm{C}$. A $\lambda/2$ is placed before the axicon to adjust the polarization of the input beam to the nonlinear crystal. The dichroic mirror, S, separates the fundamental from the second harmonic.

\begin{figure}
\centering%\sidecaption
\includegraphics*[width=\columnwidth]{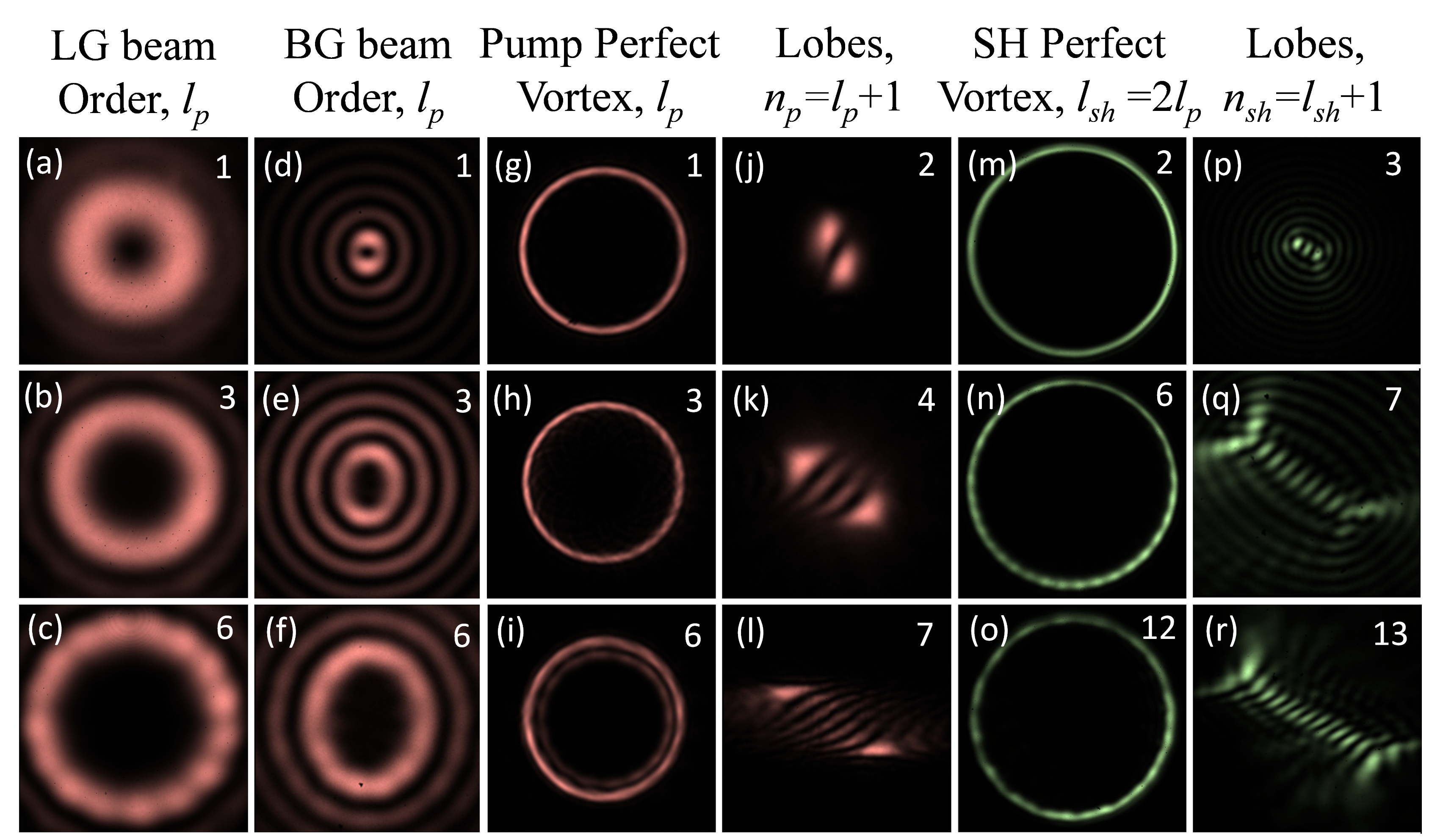}
\caption{Spatial intensity distribution of the beams recorded at different positions along beam propagation. (a-c) normal vortex pump beams of orders 1, 3 and 6 recorded before the axicon, corresponding (d-f) Bessel-Gauss beam recorded at 13 cm after the axicon, and (g-i) perfect vortices at the Fourier plane of lens L1. (j-l) characteristic lobe structure of the pump perfect vortices. (m-o) far field intensity distribution of the SH perfect vortices of orders 2, 6 and 12 and (p-r) corresponding lobe structures.}
\label{Fig2}
\end{figure}

      We have recorded the spatial intensity distribution of the pump beam and frequency-doubled green beam using a CCD-based beam profiler (SP-620U, Ophir) at different position along the propagation direction with the results shown in Fig. \ref{Fig2}.  First column, (a)-(c) of Fig. \ref{Fig2} shows the intensity profile of the pump vortices of order, $l_p =1$, 3 and 6 respectively measured before the axicon. As expected, the annular ring radius of the vortices in the first column increases with the order. The axicon converts the vortices (LG beams) into BG beams of same order, however, the BG beams maintain its structure over a distance, $z_{max}=w_{LG}/(n-1)\alpha$, where $w_{LG}$ is the beam radius of the LG beam before the axicon and $n$ and $\alpha$   are the refractive index and base angle of the axicon respectively. In our experiment, the $z_{max}$ for input vortex order $l_p=1$ is measured to be 16 cm. The size of the BG beam of order 1, 3 and 6 as shown in second column of Fig. \ref{Fig2} (d)-(f) respectively, measured at a distance $z_{max} /2$ from the axicon, increases with the order of input LG beam. Depending upon the order of the input vortices, the Fourier transforming lens L1 is placed at a distance of $\sim15.5$ cm from the axicon produces perfect vortices at the Fourier plane. The third column, (g)-(i) of Fig. \ref{Fig2}, shows the annular intensity distribution of the perfect vortices of orders $l_p =1$, 3 and 6 respectively measured at the crystal plane. From the intensity profiles of third column, it is evident that the annular ring radius of the perfect vortices are independent to their order. Given the intensity distribution of the perfect vortices, it is difficult to use interferometric technique to determine its topological charge (order). Therefore, we used tilted-lens technique \cite{Tilted_Lens} where, the perfect vortex of order \textit{l} while passing through the titled plano-convex lens (tilted about y axis at an angle $\sim6^{\circ}$) splits into $n= |l|+1$   number of bright lobes at the focal plane of the lens. From the number of bright lobes as shown in fourth column, (j)-(l) of Fig. \ref{Fig2}, it is evident that the pump perfect vortices have orders $l_p=1$, 3 and 6 respectively. Fifth column, (m)-(o) of Fig. \ref{Fig2}, represents the intensity distribution of the frequency-doubled perfect vortices recorded at the far field (distance $>2$ m away from the crystal). Using tilted-lens technique we have confirmed the orders of the SH vortices as shown in the sixth column, (p)-(r) of Fig. \ref{Fig2}, to be 2, 6 and 12 respectively, twice the order of the pump vortices. Such observation validates the angular momentum conservation in frequency-doubling process of perfect vortices. The independence of annular ring radius of the second-harmonic (SH) vortices shown in fifth column, on the orders of the pump vortices, proves nonlinear generation of perfect vortices at green wavelength. Unlike disintegration of higher order vortices in birefringent crystal \cite{6}, use of quasi-phase-matching enables generation of high quality perfect vortices (see fifth column of Fig. \ref{Fig2}) even at higher orders.   

\begin{figure}
\centering%\sidecaption
\includegraphics*[width=\columnwidth]{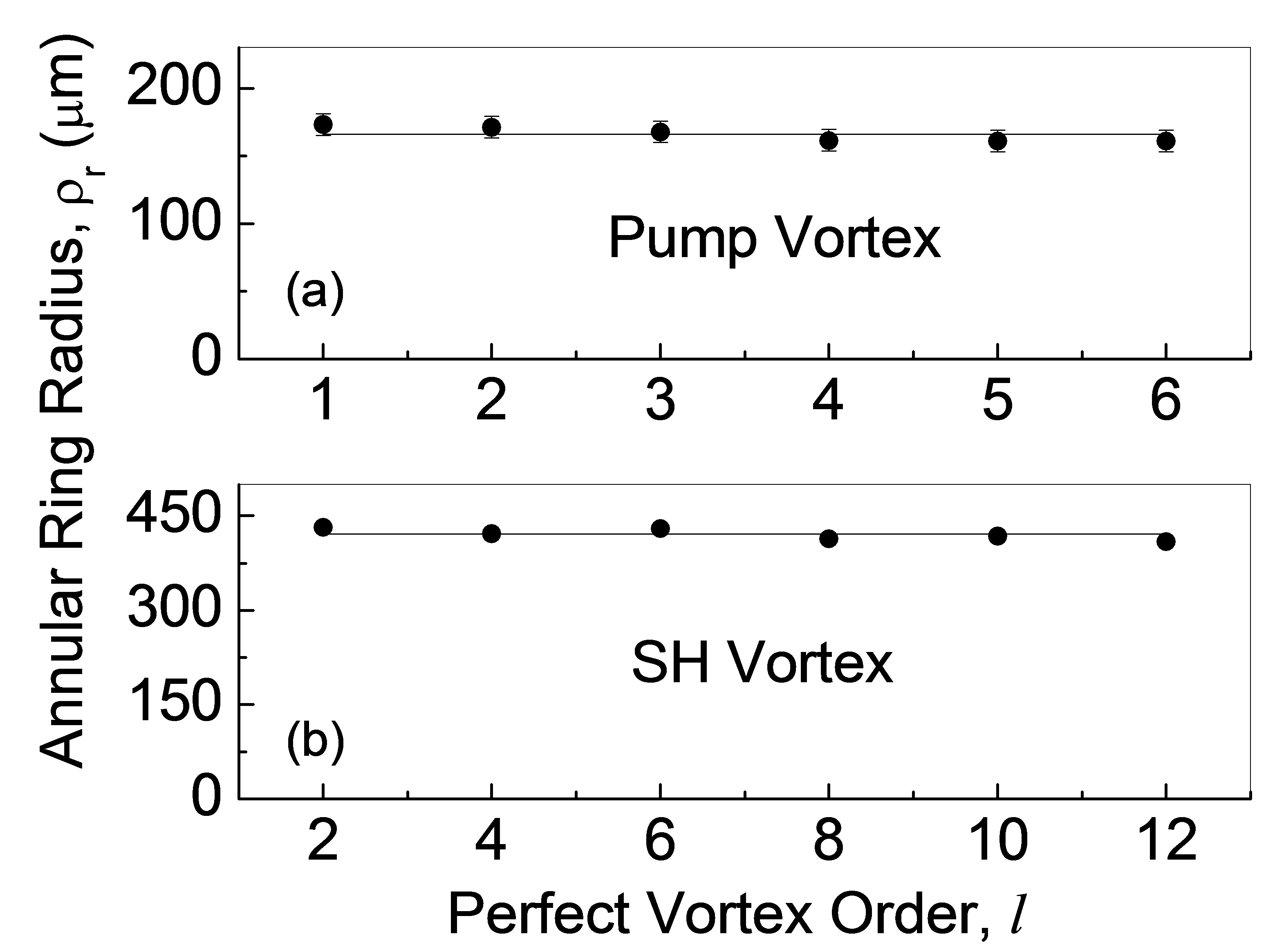}
\caption{Variation in ring radius of the annular intensity distribution of perfect vortices with its order at (a) pump and (b) SH wavelengths. The solid lines are linear fit to the experimental data. }
\label{Fig3}
\end{figure}

            We have measured the annular ring radius of the pump and SH vortices for all orders with the results shown in Fig. \ref{Fig3}. The perfect vortices at fundamental wavelength is produced at the Fourier plane of the lens L1 with measured annual ring radius of  $\sim650~\mu$m. However, to avoid mechanical constrain in accessing these perfect vortices and also to tighly focus the perfect vortices for efficient nonlinear interaction, we have imaged the pump perfect vortices using lenses L2 and L3 in $2f_2-2f_3$ configuration (magnification factor of 0.25) at the crystal plane situated at a distance 500 mm away from the Fourier plane of lens L1. As evident from the Fig. \ref{Fig3} (a), the pump perfect vortices have annular ring radius of $\rho_r^p=166\pm 6~ \mu $m for orders, $l_p=1-6$. The error in the ring radius is comparable to the pixel size (4.46 $\mu$m) of the CCD camera used to record the vortices. Pumping the nonlinear crystal with the perfect vortices, the generated SH beam is imaged at the far field ($>2$ m away from the crystal) using a lens of focal length $f=750$ mm. The variation of annular ring radius of the SH vortices with its order is shown in Fig. \ref{Fig3}(b). As evident from Fig. \ref{Fig3}(b), the SH vortices have annular ring radius $\rho_r^{sh}=420\pm 12 ~\mu $m for the orders, $l_{sh}=2-12$ confirming nonlinear generation of perfect vortices at green wavelength. The small variation in the radius of the SH vortices can be attributed mainly to the exact positioning of the CCD camera in the image plane.     
            
\begin{figure}
\centering%\sidecaption
\includegraphics*[width=\columnwidth]{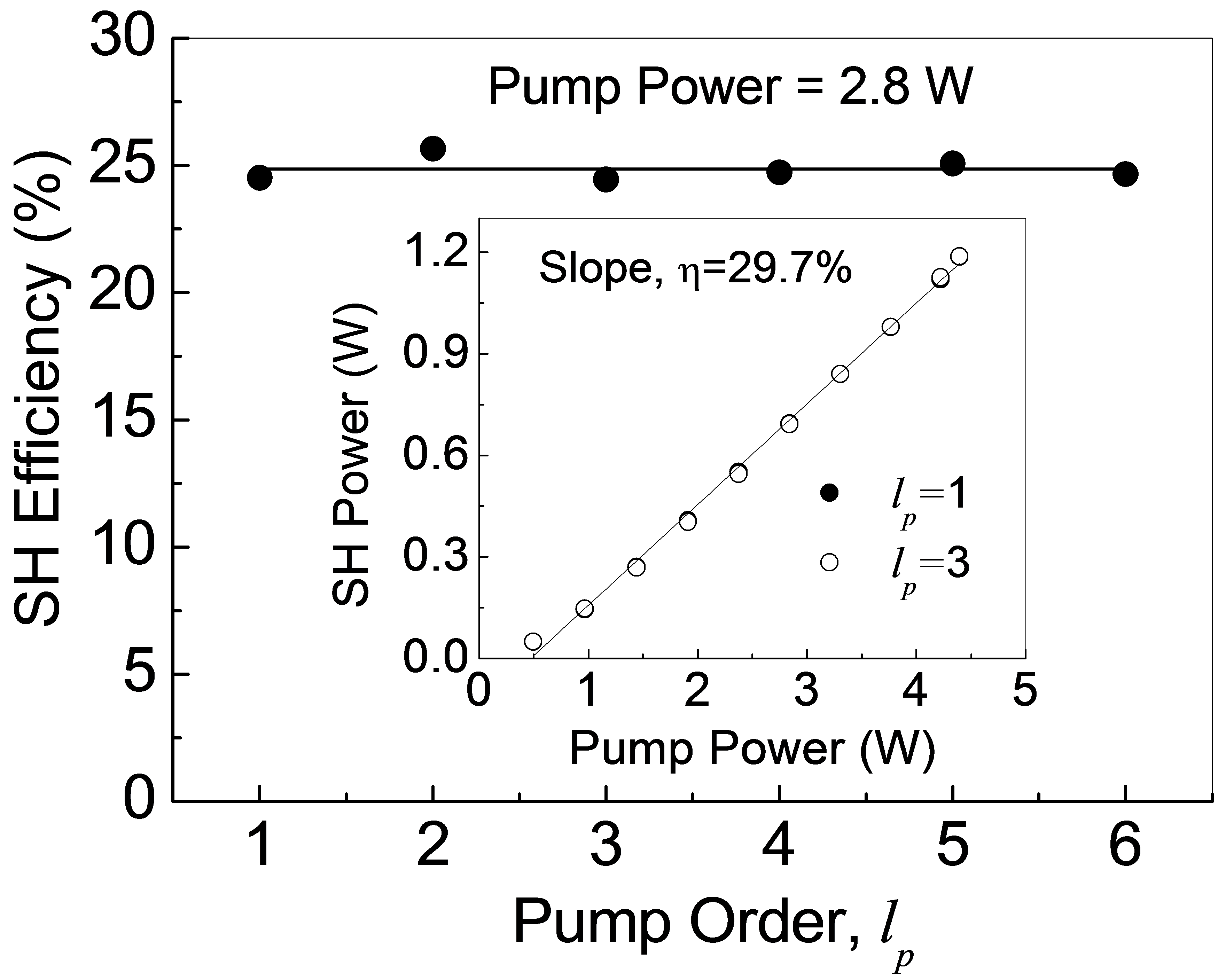}
\caption{Dependence of vortex SH efficiency with the order of the pump vortex. (Inset) variation of SH vortex power with the pump vortex power for two different orders, $l_p = 1$ and 3. Lines are linear fit to the experimental data. }
\label{Fig4}
\end{figure}

                  From Eqn. (\ref{Intensity}) it is evident that the intensity of the perfect vortex is independent of its order. Therefore, the SH efficiency which is proportional to the intensity of the input beam should be constant with the order of the input vortcies. To verify the order independent SH efficiency of the ultrafast, high power perfect vortices, we pumped the nonlinear crystal with perfect vortices of power $\sim2.8$ W and measured the SH power for different vortex orders with the results shown in Fig. \ref{Fig4}. Although the fiber laser can produce output power up to 5 W, due to losses in the vortex –doubler setup, the higher order vortices ($l_p>3$) have maximum power of $\sim2.8$ W.  As evident from Fig. \ref{Fig4}, the perfect vortices have single-pass SH efficiency of $\sim25$\% for all orders $l_p =1$ to 6. We have also measured the variation of SH power with pump power of the perfect vortices of orders, $l_p=1$ and 3 with the results shown in the inset of Fig. \ref{Fig4}. For both orders, as evident from the inset of Fig. \ref{Fig4}, the SH power increases linearly with the pump power at same slope efficiency of $\eta\sim29.7$\%. The SH perfect vortices have maximum output power of 1.2 W for the pump power of 4.4 W resulting a maximum single-pass vortex frequency-doubling efficiency of $\sim27$\%. This can be considered as the highest single-pass SHG efficiency of optical beam other than Gaussian beam.  

\begin{figure}
\centering%\sidecaption
\includegraphics*[width=\columnwidth]{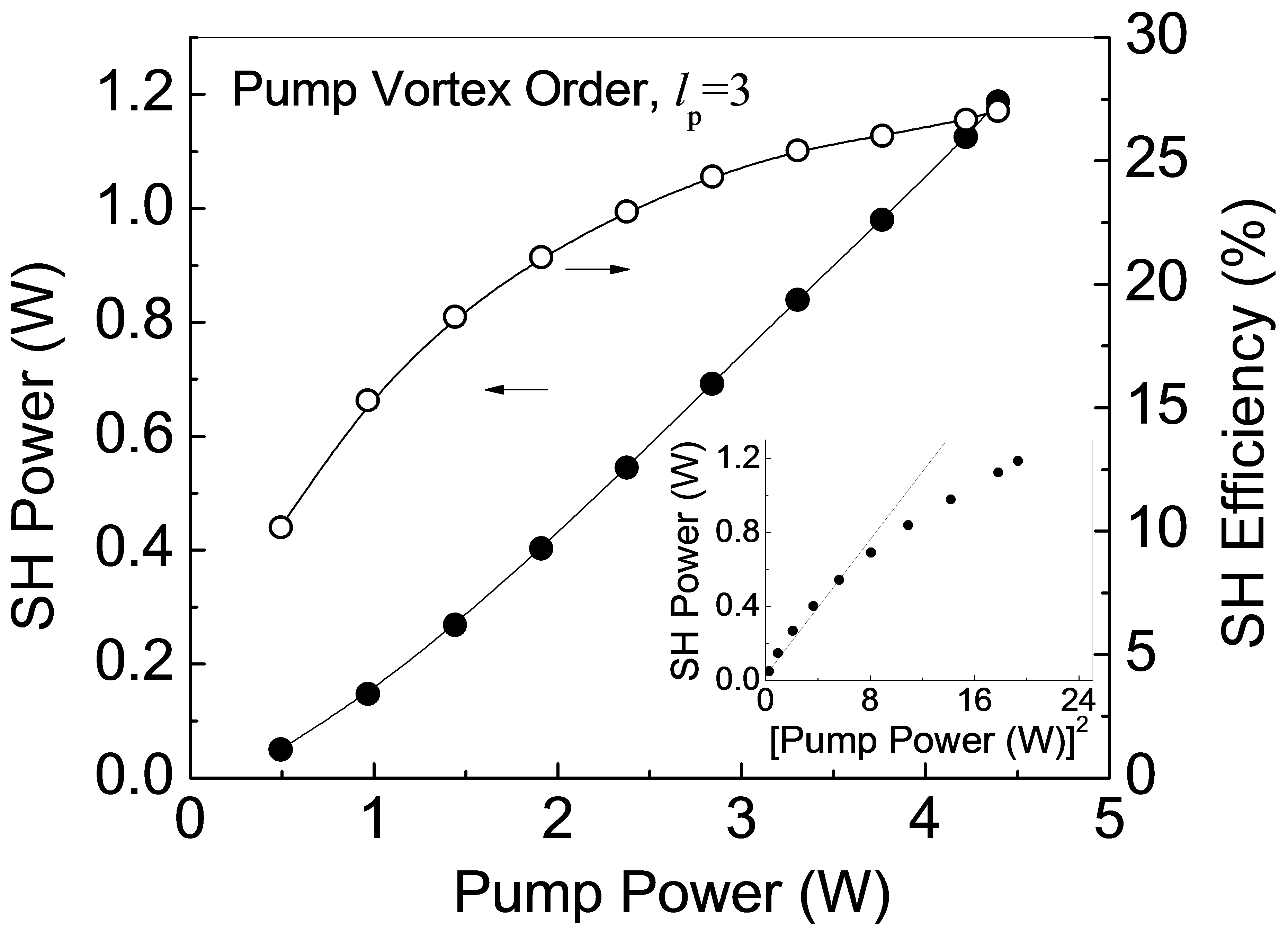}
\caption{Variation of SH vortex power and efficiency as function of pump vortex power. (Inset) Dependence of SH power with the square of the pump power. Lines are guide to eyes. }
\label{Fig5}
\end{figure}

                        Unlike quadratic dependence of SH power to the pump powers, the linear increase of SH vortex power with the pump vortex power (see inset of Fig. \ref{Fig4}), clearly indicate the saturation effect in the single-pass SH efficiency. To get further insight of the saturation effect, we have investigated the power scaling characteristics of the green perfect vortex source. Using the pump vortex of order $l_p=3$ with annular ring radius of  $\sim166~\mu$m, we have measured the SH power as a function of pump power. The results are shown in Fig. \ref{Fig5}. As evident from Fig. \ref{Fig5}, at lower pump power ($<2.8$ W), the SH power and efficiency show respectively quadratic and linearly dependence to the pump power. However, at pump power $>2.8$ W, the SH power increases linearly with the pump power and the SH conversion efficiency remains almost constant in the range of 25-27\% clearly indicating the saturation effect in the vortex SH process. The deviation of SH power from its linear dependence with the square of the pump power as shown in the inset of Fig. \ref{Fig5}, confirms the saturation effect in the vortex SHG process. Such saturation effect can be attributed to the high nonlinear parametric gain arising from high nonlinear coefficient and long interaction of the MgO:CLN crystal, and also due to the high peak power of the ultrafast pump pulses. While one can expect higher SH vortex power ($>1.2$ W) for pump vortices of smaller annular ring radius and or higher pump power $>4.4$ W, however, due to low damage threshold of the PPLN crystal \cite{PPLN_Damage} especially at the green wavelengths, we have observed crystal damage for pump power beyond 4.4 W and pump vortex radius below 166 $\mu$m. 

\begin{figure}
\centering%\sidecaption
\includegraphics*[width=\columnwidth]{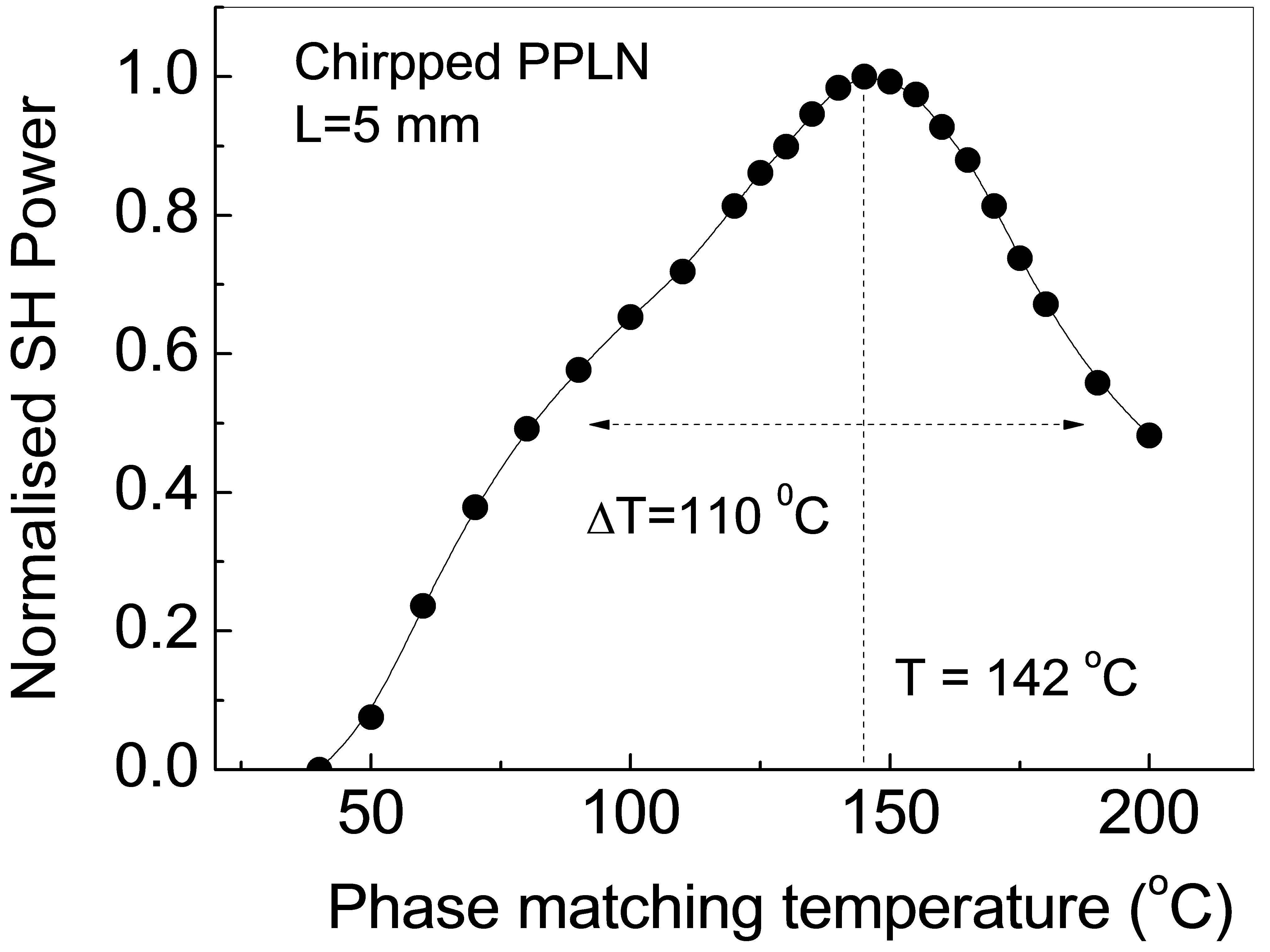}
\caption{Dependence of SH power on crystal temperature while pumped with perfect vortex of order $l_p =1$.}
\label{Fig6}
\end{figure}

                 To study the temperature dependent phase-matching characteristics of the MgO:CLN crystal, we pumped the crystal with perfect vortex ($l_p=3$) of power 1 W and measured the SH power while adjusting the crystal temperature with the results shown in Fig. \ref{Fig6}. As evident from Fig. \ref{Fig6}, the SH power increases with the increase of crystal temperature from 50 to $200^{\circ}\mathrm{C}$ with highest SH power at $T=142^{\circ}\mathrm{C}$ and a measured temperature acceptance bandwidth (full-width at half-maxima, FWHM) of $\Delta T=110^{\circ}\mathrm{C}$. As a result, the power of the SH vortices is insensitive to the fluctuation to the ambient temperature and also the instability of the temperature oven used to maintain the crystal temperature. Using a CCD based spectrometer and an intensity auto-correlator we have measured the spectral and temporal width (FWHM) of the SH vortex of all orders to be 1.9 nm centered at 530 nm and 507 fs respectively resulting a time-bandwidth product of 1.02. Similar to the Ref. \cite{6}, here we did not observe any variation in the spectral and temporal width of the SH vortices with the order.        
                 
                     In conclusion, we have experimentally demonstrated the efficient nonlinear generation of high power, higher order, ultrafast perfect vortices at the green with output power $>1.2$ W and vortex order up to $l_{sh}=12$  at single-pass conversion efficiency of 27\%. This is the highest efficiency in the single-pass SHG of any structured beam. Similar scheme can be used to generate higher order high efficient vortices at other wavelengths.

\end{document}